\documentclass[sw]{agu2001}
\usepackage{graphicx}
\authorrunninghead{MICHALEK ET AL}
\titlerunninghead{Properties and geoeffectiveness of halo CMEs}
\authoraddr{G. Michale, Astronomical Obsrvatory of Jagiellonian
University, Orla 171, Krakow, Poland. (michalek@oa.uj.edu.pl)}
\begin{document}
\title{ Properties and geoeffectiveness of halo CMEs}


\author{ G. Michalek}
\affil {Astronomical Observatory of Jagiellonian University,
Cracow, Poland}
 \author{N. Gopalswamy}
 \affil{NASA Goddard Space
Flight Center, Greenbelt, MD 20771, USA}
 \author{ A. Lara}
 \affil{Instituto de
Geof\'{\i}sica, UNAM, M\'exico}
\author {S. Yashiro}
\affil{ Center for Solar and Space Weather, Catholic University of
America}

 \begin{abstract}
  Halo coronal mass ejections (HCMEs) originating from
regions close to the center of the Sun are likely to be
geoeffective. Assuming that the shape of HCMEs  is a cone and they
propagate with constant angular widths and velocities, at least in
their early phase, we have developed a technique (Michalek et al.
2003) which allowed us to obtain the space speed, width and source
location. We apply this technique to obtain the parameters of all
full HCMEs observed by the Solar and Heliospheric Observatory (SOHO)
mission's Large Angle and Spectrometric Coronagraph (LASCO)
experiment until the end of 2002. Using this data we examine which
parameters determine the geoeffectiveness of HCMEs. We show that in
the considered period of time only fast halo CMEs (with the space
velocities higher than $\sim 1000{km\over s}$ and originating from
the western hemisphere close to the solar center could cause the
severe geomagnetic storms. We illustrate how the HCME  parameters
can be used for space weather forecast. It is also demonstrated that
the strength of a geomagnetic storm does not depend on the
determined width of HCMEs. This means that HCMEs do not have to be
very large  to cause major geomagnetic storms.
\end{abstract}

\begin{article}
\section{Introduction}
 Coronal mass
ejections (CMEs)  originating from regions close to the central
meridian of the Sun and directed toward  Earth cause the most severe
geomagnetic storms (Gosling, 1993; Kahler, 1992; Webb et al., 2001).
Many of these Earth-directed CMEs  appear as an enhancement
surrounding the  occulting disk of coronagraphs. We call them halo
CMEs (Howard et al. 1982).  The measured properties of CMEs include
their occurrence rate, direction of propagation in the plane of the
sky, angular width, and speed (e.g. Kahler, 1992; Webb, 2000; St.
Cyr et al., 2000, Gopalswamy et al., 2003a; Gopalswamy, 2004;
Yashiro et al., 2004). It is well known  that the geoffective CMEs
originate mostly within a latitude $\pm30^o$ (Gopalswamy et al.,
2000a, 2001; Webb et al., 2000, 2001; Wang et al., 2002; Zhang et
al. 2003). Srivastava and Venkatakrishan (2002) showed that the
initial speed of the CMEs is correlated with the $D_{ST}$ index
strength of the geomagnetic storm, although their conclusion was
based only on the study of four events. This tendency was also
suggested earlier by Gosling et al. (1990) and Tsurutani $\&$
Gonzalez (1998). On the other hand, Zhang et al. (2003) demonstrated
that both slow and fast HCMEs can cause major geomagnetic
disturbances. They  showed that geoeffective CMEs are more likely to
originate from the western hemisphere than from the eastern
hemisphere. They also demonstrated a lack of correlation between the
size of X-ray associated with a given CME and the importance of
geomagnetic storms. Unfortunately, these studies were based on the
sky plane speeds of CMEs without consideration of the projection
effects. The parameters describing properties of CMEs, especially
for HCMEs, are affected by projection effects (Gopalswamy et al.,
2000b). Assuming that the shape of HCMEs is a cone and they
propagate with constant angular widths and speeds, at least in their
early phase of propagation, we have developed a technique (Michalek
et al. 2003) which allows us to determine the following parameters:
the linear distance $r$ of source location measured from the solar
disk center,  the angular distance $\gamma$ of source location
measured from the plane of sky,  the angular width $\alpha$ (cone
angle =$0.5\alpha$) and  the space velocity $V$ of a given HCME. A
similar cone model was used recently by Xie et~al.~(2004) to
determine the angular width and orientation of HCMEs.

The present paper is divided into two parts. First, in the Section~2
we applied the cone model (Michalek et al. 2003)  to obtain the
space parameters of all HCMEs observed by the Solar and Heliospheric
Observatory (SOHO) mission's Large Angle and Spectrometric
Coronagraph (LASCO) until the end of 2002. In the Subsection~2.2 a
short statistical analysis, based on the derived parameters, of
HCMEs is presented  (Fig.~1~-~Fig.~4). In the Section~3,  we use
these parameters to identify the most important factors determining
geoeffectiveness of HCMEs and how they could be used for space
weather forecast (Fig.~5~-~Fig.~17)
\section{Space parameters of HCMEs}
\subsection{Data} The list of HCMEs studied in this paper is shown in
Table~1.  We considered only frontside full (type F) and asymmetric
(type A) HCMEs (Gopalswamy et al. 2003b). Only these events could be
considered using the technique proposed by Michalek et al. (2003).
Full halos are the classical halo CMEs which originate from close
the disk center. Asymmetric halos are typically wide, near-limb
CMEs, which become halos late in the event. They are different from
partial halos. The partial halos never appear around the entire
oculting disk, even in LASOC/C3 observations (their width is $<
360^o$). Only frontside events could be potentially geoeffective.
 The
 first four columns of Table~1 are from the SOHO/LASCO catalog (date, time of first appearance
 in the coronagraph field of view,
 projected speed and position angle of the fastest part of the HCME).
 Details  about  the SOHO/LASCO catalog
 and the method of measurements are described  by Yashiro et
 al. (2004).
 Parameters $r,
\gamma, \alpha,$ and $V$, estimated from the cone model (Michalek
et al. 2003), are shown in columns  (5), (6), (7), and (8),
respectively. It is important to note that for some events, the
space
 velocity determined by
this technique could be smaller than the projected speeds reported
in the LASCO catalog. This is because  the Michalek et al. (2003)
technique applies only to the beginning phase of CMEs, whereas the
CME catalog gives average speed within the LASCO's  field of view.
   The model also cannot  estimate the  parameters
for  symmetric HCMEs originating very close to the disk center and
for limb events appearing as halos on account of deflections of
preexisting coronal structures.   In column (9) the source locations
of the associated H-flares are given. The associated flares were
determined using two restrictions. They should originate in the same
part of solar disk and set up in the same time as respective CMEs
(limit time is about half an hour). To be sure that our
determination is correct we checked together EIT and LASCO movies
also. It is important  to note that localization of solar flares
might be slightly shifted with respect to origin of CMEs.   This
might affect some figures and presented correlations. By examining
the solar wind plasma data from Solar Wind Experiment (Wind/SWE,
http://web.mit.edu/space/www/wind/)  and interplanetary magnetic
field data (from Magnetic Field Investigation,
http://lepmfi.gsfc.nasa.gov/mfi), we identified, when possible, the
associated interplanetary CMEs (ICMEs). The changes of geomagnetic
indices $D_{ST}$ and $Ap$ caused by these ICMEs are presented in
columns 10 and 11, respectively. The last two columns give the
maximum value of magnitude ($B$) and southward component ($B_Z$) of
magnetic field in the ICME. We considered 144 frontside HCMEs
(FHCMEs) recorded by the LASCO coronagraphs until the end of 2002.
For 101(70$\%$) of them we were able to determine the required
parameters ($r, \gamma, \alpha,$ and  $V$).
  The  events that could not be measured were
mostly too faint to get  height-time plots at the opposite sides
of the occulting disk. Only a few  (16) were symmetric for which
we could not obtain the HCME parameters.

\subsection{ Statistical analysis}

 \subsubsection{ The space velocities  of FHCMEs}

The properties of halo CMEs observed by SOHO/LASCO have been
described in a number of papers (Gopalswamy et al., 2003a;
Gopalswamy, 2004; Yashiro et al., 2004). Here we describe the
properties of FHCMEs measured
 according to Michalek et al. (2003).
 Fig.~1 shows the distribution of the space velocities
 ($V$)
 of FHCMEs during  the ascending (1996-1999) and  maximum
 phases of solar activity (2000-2002) as well as for the whole period
 (1996-2002).
  It was noted before, e.g., by Webb et al. (1999), Gopalswamy (2004, see Fig~1.13,~1.14) and
  Yashiro et al. (2004), that
 HCMEs are much faster and more energetic than typical CMEs.
 Our results also confirm this. The average speed of the HCMEs
 is $1300~km/s$ (about $25\%$  larger than that for HCMEs from SOHO/LASCO
catalog, Yashiro et al. (2004)). The difference, between average
speeds received in the present paper and by Yashiro et al. (2004),
is likely to be due to the fact that we are using corrected speeds
while Yashiro et al. (2004) used sky-plane speeds. We use a smaller
number of events in the statistic. From the histograms in Figure~1,
it is evident that velocities of HCMEs increase  significantly
following the solar activity cycle as for all CMEs (Yashiro et al.,
2004). During the maximum of solar activity the FHCMEs have, on the
average, velocities about $40\%$ higher than the average velocities
during the minimum of solar activity. The speed of the slowest event
is $189~km/s$ while the speed of the fastest one is $2655~km/s$.

  In
 Fig.~2, we present the sky-plane speeds against the corrected
(space) speeds.  The solid line represents the linear fit to the
data points. The inclination of the linear fit demonstrates that
the projection effect increases slightly
 with the speed of CMEs.  It is clear that the projection
effect is important, and on average the corrected speeds are
$25\%$ higher than the velocities measured in the plane of sky.
This was also anticipated based on other considerations
 (Gopalswamy et al., 2001). It
is important to note  that both sky-plane and corrected speeds are
determined at the same distance ($2R_{\odot}$) from the disk
center.

\subsubsection{Widths of FHCMEs}
 Fig.~3 shows  the distribution of the estimated widths
($\alpha$)
 of FHCMEs during the ascending  (1996-1999) and maximum phases of
 solar activity
 (2000-2002)  as well as for the whole period (1996-2002).
 The average width of HCMEs is  $120^o$ (more than twice
 the average value obtained from the SOHO/LASCO catalog, Yashiro
et al., 2004). The average width of HCMEs does not change
significantly with solar activity, except for   a small increase
during  the maximum of solar activity. The most narrow HCME has a
width of $39^o$ and the widest one has $\alpha$  as large as
$168^o$.

\subsubsection{Source locations of FHCMEs}
 Fig.~4 presents the distribution of source
location ($\gamma$)
 of FHCMEs during  the ascending (1996-1999) and maximum phases of
 solar activity
  (2000-2002) and for the whole period (1996-2002).
  FHCMEs with $\gamma$ close to
$0^o$ originate near to the  solar limb while events  with
$\gamma$ close to $90^o$ originate from the disk center region.
Fig.~4 shows that the FHCMEs originate close to the Sun center
 with a maximum of distribution around $\gamma=62^o$. The
 distribution of source location does not depend on the period of
 solar activity. We have to note that these distributions are slightly
 biased due to the fact that we neglected 16 symmetric
 FHCMEs (these CMEs cannot  be measured using the cone model).
 They originate very close to the disk center and should slightly
 increase the average value of $\gamma$.

\section{Geoeffectiveness of  FHCMEs.}
 Having defined the parameters
describing FHCMEs, we now explore which of theses parameters
determine the strength of geomagnetic disturbances. In situ
counterparts of frontside HCMEs can be recognized in the magnetic
field and plasma measurements as ejecta (EJs) or magnetic clouds
(MCs). Magnetic clouds can be identified by the  following
characteristic properties: (1) the magnetic field strength is
higher than the average;  (2) the proton temperature is lower than
the average; (3) and the magnetic field direction rotates smoothly
(Burlaga 1988, 2002, 2003a,b; Lepping et al., 1990).  In the
present paper we refer to both MCs and EJs as interplanetary CMEs
(ICMEs). The presence of these signatures changes from one ICME to
an other. By examining the solar wind plasma data  we identified,
when possible, ICMEs. These ICMEs could be responsible for
geomagnetic disturbances.
  The
strength of geomagnetic storms is described  by two indices Ap
(which measures the general level of geomagnetic activity over the
globe) and $D_{ST}$ (which is obtained using  magnetometer data from
stations near the equator). The maximum values of $D_{ST}$ and $Ap$
indices associated with the ICMEs are presented in Table~1.  We
included those events for which the $D_{ST}$ index decreased below
-25nT. We now examine the relation between the geomagnetic indices
and $V$, $\alpha$ and $\gamma$. First in the Subsection~3.1, we
consider influence of different parameters on geoeffectiveness of
FHCMEs (Fig.~5~-~Fig.~14). In the Subsection~3.2 we try to find
which FHCMEs could cause false alarms (Fig.~15~-~Fig.~17).
\subsection{Geoeffectiveness of FHCMEs}
 \subsubsection{Geoeffectiveness and
space velocity ($V$) of FHCMEs.}
 Fig.~5 shows the  scatter plots of plane of sky speeds
versus $D_{ST}$ and $Ap$ indices.    Diamond symbols represent
events originating from the western hemisphere and cross symbols
represent events originating from the eastern hemisphere. The solid
lines are the linear fits to the data points associated with eastern
events, and the dashed lines are linear fits to data points
associated with western events. The dot-dashed vertical lines
indicate velocity limits above which HCMEs can cause geomagnetic
storms with $D_{ST}\leq -150nT$. These lines were inferred from two
events on 1 May 1998 and 2 May 1998. Upon inspection of this figure,
it is clear that the major geomagnetic storms can be generated by
slow (speeds $\approx 500km/s$) and fast HCMEs originating in the
western hemisphere. There is not a significant correlation
(correlation coefficients are $< 0.50$) between the projected speed
and geomagnetic indices. Linear and Spearman correlation
coefficients are approximately equal 0.35(0.31) for the western and
0.10(0.05) for eastern events, similar to the results of Zhang
et~al.~(2003). The situation is different when we consider the space
velocities of HCMEs. In Fig.~6, the scatter plots of $V$ versus
$D_{ST}$ and $Ap$ indices are presented. The space velocities are
larger than the plane of sky speeds and all events in the panels are
shifted to the higher velocity range, especially for the two events
on 1 May 1998 and 2 May 1998, which seem to be narrow (width
$\approx 40^o$) and three times faster than they appear in LASCO
observations (Table~1). Determination of the space velocity is
consistent with observations of ICMEs associated with these events.
Since, these CMEs  needed only $\approx 46$ hours to reach Earth
(Manoharan et al., 2004, Michalek et al., 2004), they must be very
fast.
  In  LASCO observations these CMEs appear  faint,
 suggesting that they seem to be narrow and they could be observed
 as halos when they are far from the Sun. Upon inspection of this
 figure, it is clear that only very fast events ($V\geq1100km/s$) originating in the western
hemisphere can cause the biggest geomagnetic storms ($D_{ST}\leq
-150nT$). The dot-dashed verticals lines indicate velocity limits
above which HCMEs can cause severe geomagnetic storms. We find
significant correlation (correlation coefficients are $>0.50$)
between velocity and geomagnetic indices for the western events. The
linear and Spearman correlation coefficients are 0.60(0.54) and
0.62(0.56) for $Ap$ and $D_{ST}$ indices respectively. In contrast
there is very little correlation between the space velocity and
geomagnetic indices for the eastern events. the linear and spearman
correlation coefficients are 0.16(0.04) and 0.07(0.02) for $Ap$ and
$D_{ST}$ indices, respectively. Events originating in the eastern
hemisphere are not likely to cause  major geomagnetic storms. Fig.~7
shows the distribution of the space velocities $(V)$ of FHCMEs,
which cause geomagnetic disturbance with $D_{ST}$ index lower than
$-25nT$, $-60nT$ and $-100nT$, respectively. These histograms
demonstrate again that geoeffectivenes of HCMEs depend on their
space velocities and sever geomagnetic storms with $D_{ST}<-100nT$
can be caused by fast CMEs (with $V>700km/s$) only. The results seem
to be different from these reported by Zhang et~al.~(2003) for the
plane of sky speeds. This demonstrates that conclusions based on
coronagraphic observations subjected to the projection effects could
be incorrect.

 \subsubsection{Geoeffectiveness and $\gamma$ of HCMEs.}  In  Fig.~8
 we show
 the scatter plots of $\gamma$ versus $D_{ST}$ and $Ap$ indices.
 Diamond symbols represent events originating from the western
hemisphere and cross symbols represent events originating from the
eastern hemisphere. The solid lines are the linear fits to the data
points associated with eastern events and the dashed lines are
linear fits to data points associated with western events. For the
eastern  events the correlation between $\gamma$ and geomagnetic
indices is not significant. For these events, the linear and
Spearman correlation coefficients for $Ap$ and $D_{ST}$ indices are
0.14(0.18) and 0.20(0.38), respectively. In contrast the western
events originating close to the disk center ($\gamma\geq65^o$) are
more likely to cause the biggest geomagnetic storms. For these
events correlation coefficients for $Ap$ and $D_{ST}$ indices are
0.39(0.38) and 0.35(0.42), respectively. Similar conclusions are
obtained when we consider H-alpha flare locations. In Fig.~9 we show
the scatter plots of absolute values of longitudes of H-alpha flares
associated with HCMEs versus $D_{ST}$ and $Ap$ indices. Diamond
symbols represent events originating from the western hemisphere and
cross symbols represent events originating from the eastern
hemisphere. These results  confirm previous conclusion that the
western events originating close to the disk center are more likely
to cause the biggest geomagnetic storms. We have to note that there
is one event on 04 April 2000 which originate far from the disk
center (N26W66) and cause the severe geomagnetic storm with
$D_{ST}=-288nT$.   Now the correlation between longitude and
geomagnetic indices is very poor for the western and eastern events
as well. The results are proved by histograms presented in Fig.~10.
This figure presents the distribution of the longitude of FHCMs
which cause geomagnetic disturbance with $D_{ST}$ index lower than
$-25nT$, $-60nT$ and $-100nT$. Upon inspection of the histograms, it
is clear that the goeffectiveness of CMEs depends on the longitude
of source location and that the severe geomagnetic disturbance
($D_{ST}<-100nT$) are mostly caused by the western events
originating close to the disk center. During the study  period of
time there were only two severe geomagnetic storms ($D_{ST}<-100nT$)
caused by western events originating far from the disk center. It is
important to note that the peak of the longitude distribution is
shifted to the west from the disk center.
 \subsubsection{Geoeffectiveness and angular widths
($\alpha$) of CMEs} In  Fig.~11, we have shown the  scatter plots of
$\alpha$ against $D_{ST}$ and $Ap$ indices.  Cross and diamonds
symbols are associated with the eastern and western events,
respectively.  The solid lines are the linear fits to the data
points associated with eastern events and the dashed lines are
linear fits to data points associated with western events.
 Upon
inspection of the figures it is clear that the geoeffectiveness of
CMEs depends very little on their widths. All considered correlation
coefficients are $\leq0.22$. Even severe geomagnetic storms can be
caused by both narrow and wide HCMEs.  This means that HCMEs do not
have to be very large  to cause  major geomagnetic storms.

\subsubsection{Geoeffectivness against velocities and source
localization of the FHCMEs} As we demonstrated in the previous
subsection, geoeffectiveness of FHCMEs strongly depends on their
space velocity $V$ and source location $\gamma$. These parameters
may be helpful for space weather forecast. In  Figs.~12 and 13, $Ap$
and $D_{ST}$ indices versus $\gamma$ and $V$ are shown in contour
plots using   the Kriging (Isaacs and Srivastava, 1989) procedure
for generating regular grids. The darker the shade, the higher are
the $Ap$ and $D_{ST}$ indices. Knowing the source location and space
velocity of a given HCME, we can, in a simple way, predict its
geoeffectiveness. From the inspection of the picture we see that the
strongest geomagnetic storms can occur for fast events originating
close to the disk center.

\subsubsection{Geoeffectiveness and interplanetary magnetic field
(IMF) carried by FHCMEs}

Fig.~14 shows the scatter plots of the maximum values of magnitude
($B$) and southward component ($B_Z$) of ICME magnetic field versus
$D_{ST}$ and $A_{p}$ indices. The solid lines are linear fits to the
data points. This figure clearly confirms  that the major
geomagnetic storms are generated by CMEs carrying strong magnetic
filed with significant southward component. Correlation between
these parameters and geomagnetic indices is very large and linear
coefficients are   approximately equal $0.70$ for ($B$) and ($B_Z$)
as well. Spearman correlation coefficients in this case are slightly
smaller and are approximately equal $0.60$ for  ($B$) and ($B_Z$).
Spearman correlation coefficients in this case are slightly smaller
and are approximately equal $0.60$ for  ($B$) and ($B_Z$). It is due
to the fact that they are derived from the rank of variable within
the distribution and they are not sensitive to the 4 outlying points
with $|B|>30nT$. Unfortunately, $B$ and $B_Z$ are measured in situ
and hence may not be useful for space weather forecast.

\subsection{False alarms} Previous studies (e.g. Cane et al., 2000;
St. Cyr et al., 2000; Wang et al. 2002) have shown that a large
fraction of frontside HCMEs is non-geoeffective.  St. Cyr et al.
(2000) found that only $20/40$ ($50\%$) of all frontside HCMEs
during 1996-1998 from SOHO/LASCO caused geomagnetic storms with
$K_{P}\geq 5$. Wang et al. (2002) used a larger data base (March
1997 to December 2000) showed that $59/132$ ($45\%$) of frontside
HCMEs could result in moderate to severe geomagnetic storms
($K_{P}\geq 5$) and that the majority of these events occurred
within latitude $\pm(10^{o}-30^{o})$. They also found an asymmetry
in the central meridian distance distribution. In the western
hemisphere, a geoeffective event could be expected even at
$\sim70^o$. On the eastern side, there were no geoeffective HCMEs
outside of $40^o$. We performed a similar analysis for our sample of
FHCMEs. During the study period, there were only $88/144$ FHCMEs
with geomagnetic signatures ($D_{ST}>-25nT$) at Earth which means
that only $60\%$ of FHCMEs are geoeffective. For $65/88$ ($73\%$) of
them we determined the space parameters ($r, \gamma, \alpha,$ and
$V$). If we take into account only those FHCMEs which caused
moderate to severe geomagnetic storms ($D_{ST}>-60nT$) the fraction
of geoeffective events decreased to $51/144$ ($36\%$). It is
important to recognize them because they generate "false alarms". In
our sample, there were $56/144$ ($39\%$) not geoeffective HCMEs. For
$36/56$ ($62\%$) of them we were able to determine the space
parameters ($r, \gamma, \alpha,$ and $V$). We now explore why these
FHCMEs did not cause geomagnetic disturbances. Fig.~15 presents the
distributions of longitude and the space velocities of the $36$
non-geoeffective FHCMEs. The histograms show that these events
originate from the whole solar disk and have velocities from
$100km/s$ up to $2500km/s$. The distributions do not demonstrate any
specific signatures characterizing these events. Fig.~16 shows, in
the successive panels, the distributions of: the space velocities
for FHCMEs originating close to the disk center
($|longitude|<30^o$), the space velocities for FHCMEs originating
close to the limb ($|longitude|>30^o$), the longitude for slow
FHCMEs ($V<1200km/s$) and the longitude for fast FHCME
($V>1200km/s$). Upon the inspection of the histograms (the first and
last panel in the figure) it is clear that all fast HCMEs
($V>1200km/s$) originating close to the disk center
($|longitud|<30^0$)  must be geoeffective. There is no false alarm
for such events. Slower FHCMEs ($V<1200km/s$) originating close to
the disk center do not have to be geoeffective. In the third panel
we note $20$ events originating from the disk center but not
influencing Earth. On the other hand, even very fast FHCMEs
($V>1200km/s)$ were not geoeffetive when they originated close to
the limb. In the fourth panel we have 16 fast FHCMEs originating
close to the limb without geomagnetic signatures at Earth. Fig.~17
shows the scatter plot of the space velocities versus longitude for
all FHCMEs. Diamond symbols represent goeeffective and cross symbols
non-geoeffective FHCMEs. The solid lines are linear fits for
non-geoeffective events originating from the east and west
hemisphere. For non-geoeffective eastern and western events the
linear and Spearman correlation coefficients are very large and
equal $0.88(0.87)$ and $0.79(0.74)$, respectively. Upon inspection
of the figure, it is clear that geoeffective events are faster than
the
 non-geoeffective events originating at the same
longitude. It is also clear from the strong correlation that
events originating farther form the disk center are faster than
those originating close to the disk center. Linear fits to the
non-geoeffective events  could be considered as lower limits for
the space velocities above which a given CME originating at a
given longitude could be observed as a halo event. In the vicinity
of these fits we see both geoeffective and non-geoeffective
FHCMEs. Slightly above these fits we see only geoeffective FHCMEs.
It is important to note that the inclination of linear fit to the
eastern events is steeper than that for the western events.
Eastern events must be faster to appear as halo events or to be
geoeffective than western events originating at the same angular
distance from the disk center.

Generally our results are consistent with those of previous
studies. We would like to emphasize that the geoeffectiveness of
HCMEs depends not only on source locations, but also on their
space velocity. Having both the parameters improves our ability to
forecast whether a given HCME will be geoeffective or not. Non
geoeffective events are slow or fast but originating far from disk
center. They do not affect  magnetosphere. If they are directly
ejected to Earth they are slow and disturbed before reach Earth.
If they are fast, they are ejected not directly to Earth (events
with large longitude) and they only touch magnetosphere by flanks.
Unfortunately there is very difficult to give sharp boundary
limits dividing CMEs on geoeffective and non geoeffective events.
These limits depend not only on CMEs properties but also on
condition of interplanetary medium. Approximate limits can be
obtain from Fig.~17. Of course, we appreciate that additional
parameters such as the strength and orientation of the  resulting
interplanetary CME are also expected to play a role in deciding
the geoeffectiveness. It is difficult to give sharp boundary
conditions for non geoeffective events.

\section{Summary}
 In this study we considered the geoeffectiveness of all full HCMEs
 observed by SOHO/LASCO coronagraphs from the launch in 1995 until
 the end of 2002. For $101/144$ ($70\%$) of full HCMEs we were able to find the source location, width and
space velocity using the cone model (Michalek et al., 2003). We must
be aware  that the cone model is only rough simplification of real
events. We know that not all CMEs are perfectly symmetric (Moran and
Davila, 2004; Jackson et al., 2004). Most of CMEs could be
approximate using cone model but probably for some of them this
assumption is unrealistic. Fortunately technique presented by
Michalek et al. does not demand  perfect symmetry for CMEs. This
approach requires measurements of sky-plane speeds and the moments
of the first appearance of the halo CMEs above limb at only two
opposite points. We are able to determine,  with good accuracy, the
space velocity and with of a given CME at least in the plane
symmetry crossing CMEs at these points. When a given CME could be
approximated by the cone model these derived parameters are  valid
for the entire CME. HCMEs originating very close to the disk center
(mostly within a latitude of $\pm40^o$), are very wide (the average
angular width $=120^o$) and are very fast (the average space speed
$=1291km/s$). We find significant ($40\%$) increase in the average
space velocities of HCMEs during the maximum of solar activity.
These results could suggest that the HCMEs represent a special class
of CMEs which are very wide and fast. It is important to note that
this "class" of CMEs is defined due to artificial effect caused by
coronagraphic observations. Events originating close to the disk
center (from SOHO/LASCO point of view) must be wide and fast to
appear as HCMEs in LASCO observations. This is not due to
localization on the solar disk but due to  oculting disk which not
only blocks bright photospheric  light but also eliminates some
narrow and slow events. We have to emphasize  that this effect
mostly depends on the dimension of oculting disk but in less degree
on the sensitivity of instrument. More sensitive instrument can
record some poorer events (halos and also not halos so statistic
will be similar) but could not register these events which never
appear behind occulting disk. Potentially more sensitive instrument
could register less energetic events (narrower and slower) and the
average velocities and widths (for halos and whole population of
CMEs) could be slightly lower but the main relation between the
halos and   whole population of events will be the same.
Fortunately, poor events do not cause a big concern because they are
not geoeffective. We do not expect, in the near future, any  special
programs devoted to looking for less energetic CMEs. Next scientific
mission (STEREO) will be mostly dedicated to recognize 3D structure
of CMEs.
 Such fast and wide CMEs are known to be associated with electron
and proton acceleration by driving fast mode MHD shocks (e.g., Cane
et al., 1987; Gopalswamy et al., 2002a). Using observations from
Wind spacecraft, interplanetary magnetic clouds (MC) and geomagnetic
disturbances associated to HCMEs were identified. The strength of
geomagnetic storms, described by $D_{ST}$ and $Ap$ indices, is
highly correlated with the source location and space velocity of a
given event. Only HCMEs originating in the western hemisphere, close
to the solar center and very fast (space velocity $\geq 1100km/s$)
are likely to cause major geomagnetic storms ($D_ST<-150nT$). Slow
HCMEs (space velocity $\leq 1100km/s$), even originating close to
the solar center, may not cause severe geomagnetic disturbances. We
have to note that there was one event (04 April 2000), which
originated far from disk center and produced a severe geomagnetic
storm ($D_{ST}=-288nT$).
 Probably this storm was not due to an ICME. It was caused by the sheath
 region ahead of the CME as was reported by Gopalswamy (2002b).
 We illustrated, using contour maps, how the derived HCME parameters
 can be useful for space weather forecast. We have to note that geoeffectiveness of events
does not depend on their widths.

During our study period we recognized $56/144$  ($30\%$) FHCMEs
without any geomagnetic signature at Earth.  This is significant
population of FHCMEs. To distinguish them from the geoeffective
events we considered the source locations and space velocities of
HCMEs. When both the parameters are available, it becomes easier to
assess the geoeffectiveness of HCMEs. We may say that fast FHCMs
($V>1200km/s$) originating close to the disk center
($|longitude|<30^o$) must be geoeffective. For such events there
were no false alarms. But, even very fast events originating far
from the disk center can be non-geoeffective.

\begin{acknowledgements}
 This work was done when GM visited the Center
 for Solar Physics and Space Weather, The
Catholic University of America in Washington.\\
Work done by Grzegorz Michalek was partly supported by {\it
Komitet Bada\'{n} Naukowych} through the grant
PB~0357/P04/2003/25. Part of this research was also supported by NASA/LWS and NSF/SHINE programs.
\end{acknowledgements}
{}
\newpage

\begin{figure*}
\vspace{9cm} \includegraphics{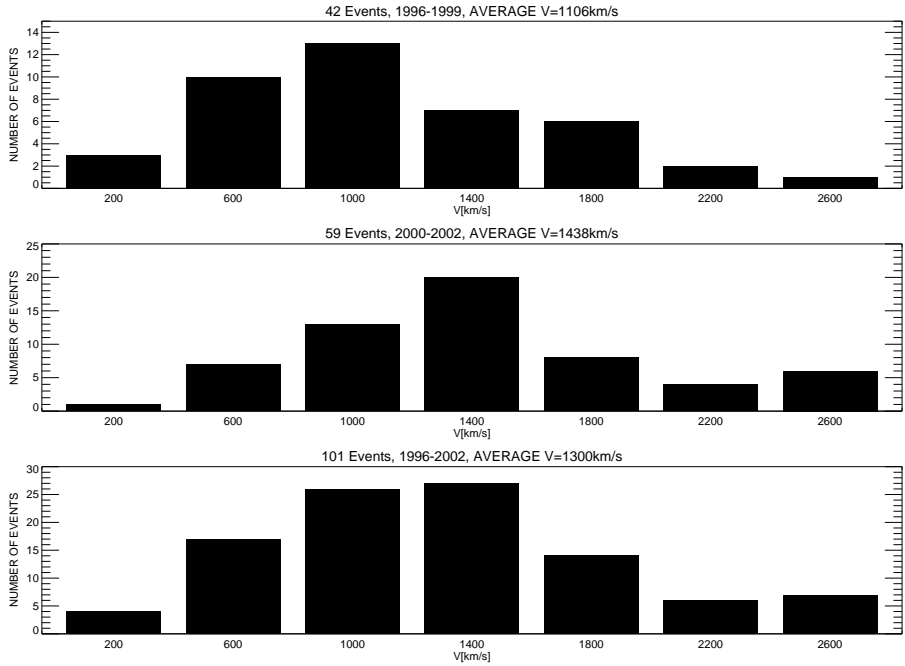} \vspace{0mm} \caption{The histograms
showing the distribution of space velocities
 ($V$)
 of HCMEs during the ascending  (1996-1999) and maximum
 phases of solar activity (2000-2002) and for the whole considered period (1996-2002).
From the histograms, it is evident that velocities of HCMEs increase
significantly following the solar activity cycle. }
\end{figure*}

\begin{figure*}
\vspace{9cm} \includegraphics{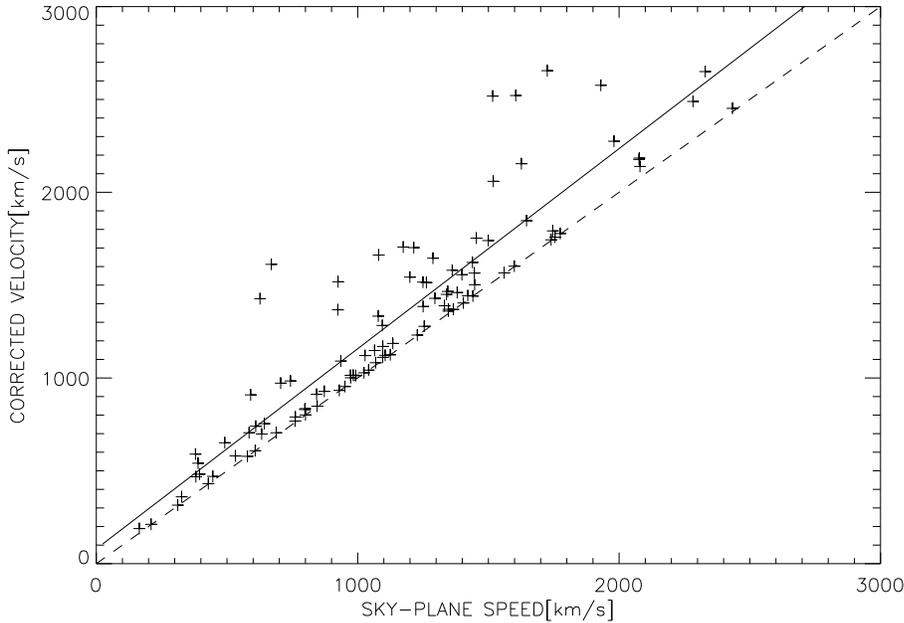} \vspace{0mm}\caption{The plane of  sky speed versus
the corrected (space) speed of HCMEs. The solid line shows the
linear fit to data points. The inclination of the linear fit
demonstrates that the projection effect increases slightly
 with the speed of CMEs.}
\end{figure*}
\newpage


\begin{figure*}
\vspace{9cm} \includegraphics{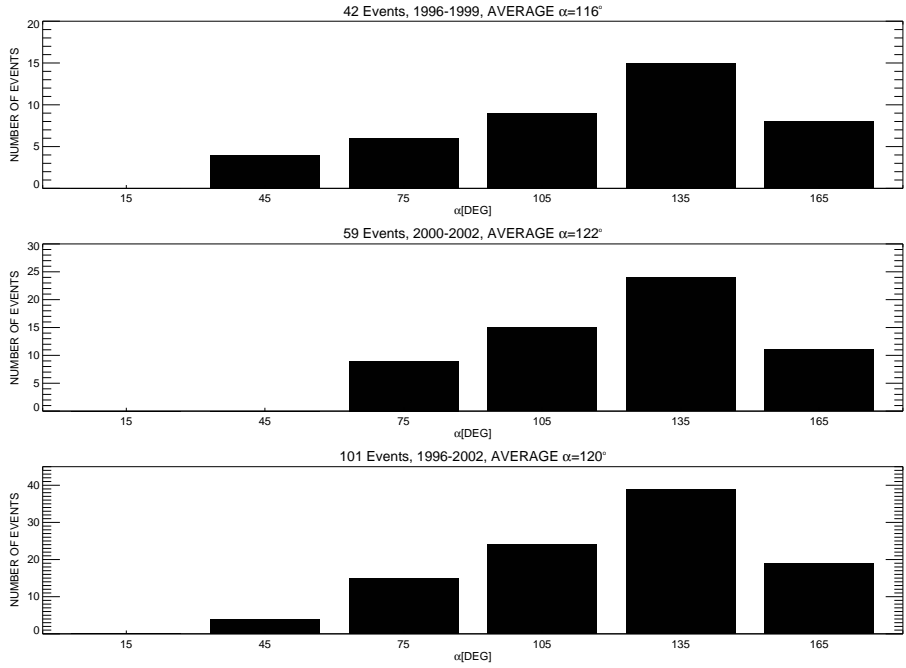} \vspace{0mm} \caption{The histograms
showing the distribution of width ($\alpha$) of HCMEs during the
ascending  (1996-1999) and  maximum
 phases of solar activity (2000-2002) and for the whole considered period (1996-2002).
The average width of HCMEs does not change significantly with
solar activity}
\end{figure*}

\newpage
\begin{figure*}
\vspace{19cm} \includegraphics{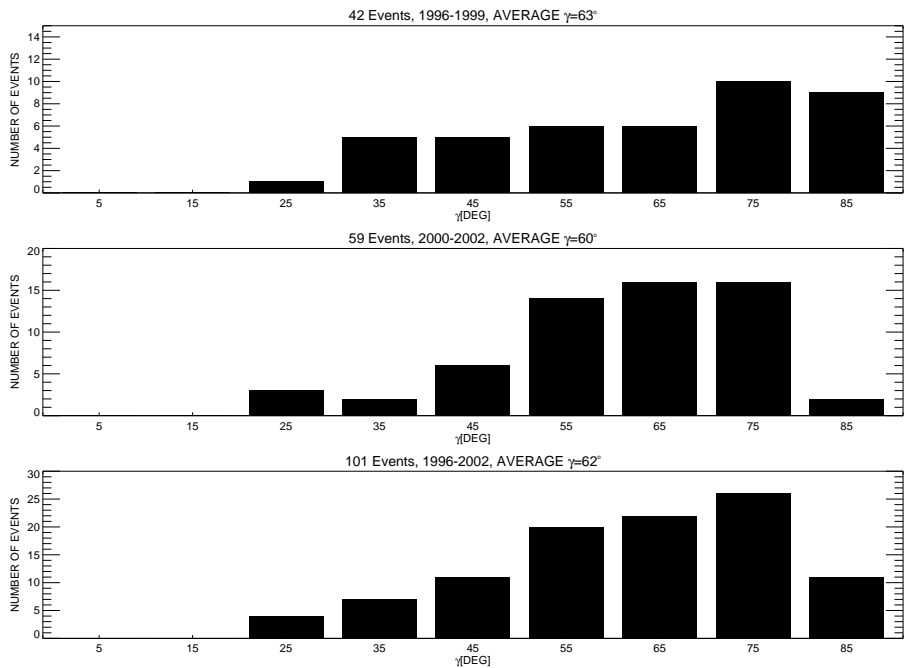} \vspace{0mm} \caption{The histogram
showing distribution of source location ($\gamma$) of HCMEs during
 the ascending  (1996-1999) andmaximum
 phases of solar activity (2000-2002) and for the whole  considered period (1996-2002).
 Histograms shows that the FHCMEs originate close to the Sun center
 with a maximum of distribution around $\gamma=62^o$. The
 distribution of source location does not depend on the period of
solar activity.}
\end{figure*}

\begin{figure*}

\vspace{9cm} \includegraphics{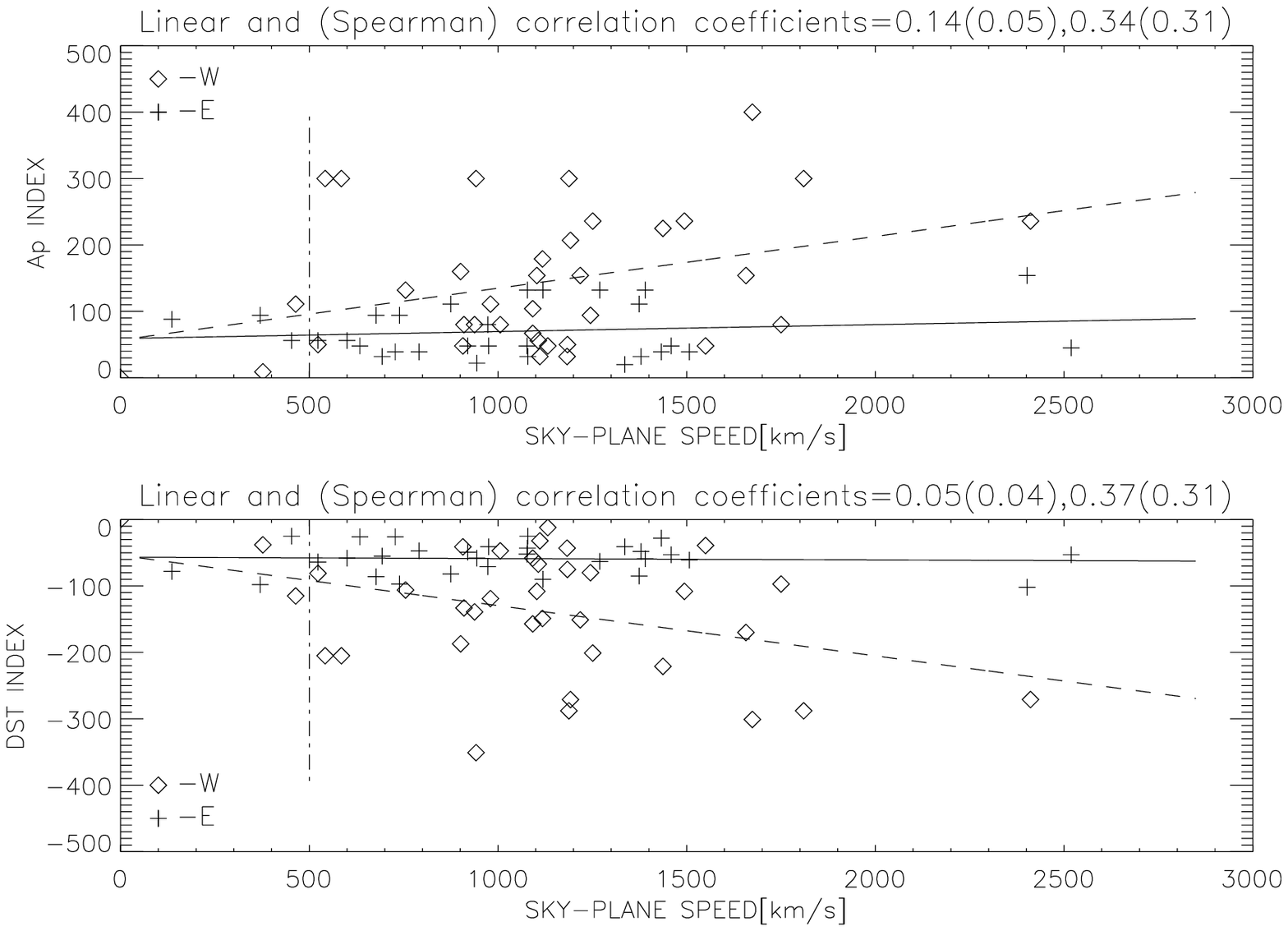} \vspace{0mm}\caption{The scatter plots of the
sky-plane speed versus $Ap$ and $D_{ST}$ indices. Diamond symbols
represent events originating from the western hemisphere and cross
symbols represent events originating from the eastern hemisphere.
The solid lines are the linear fits to  data points associated with
eastern events, and the dashed lines are linear fits to data points
associated with western events. The dot-dashed vertical lines
indicate velocity limits above which HCMEs can cause severe
($D_{ST}\leq -150nT$) geomagnetic storms. Upon inspection of this
figure, it is clear that the major geomagnetic storms can be
generated by slow (speeds $\approx 500km/s$) and fast HCMEs
originating in the western hemisphere.}
\end{figure*}

\begin{figure*}

\vspace{9cm} \includegraphics{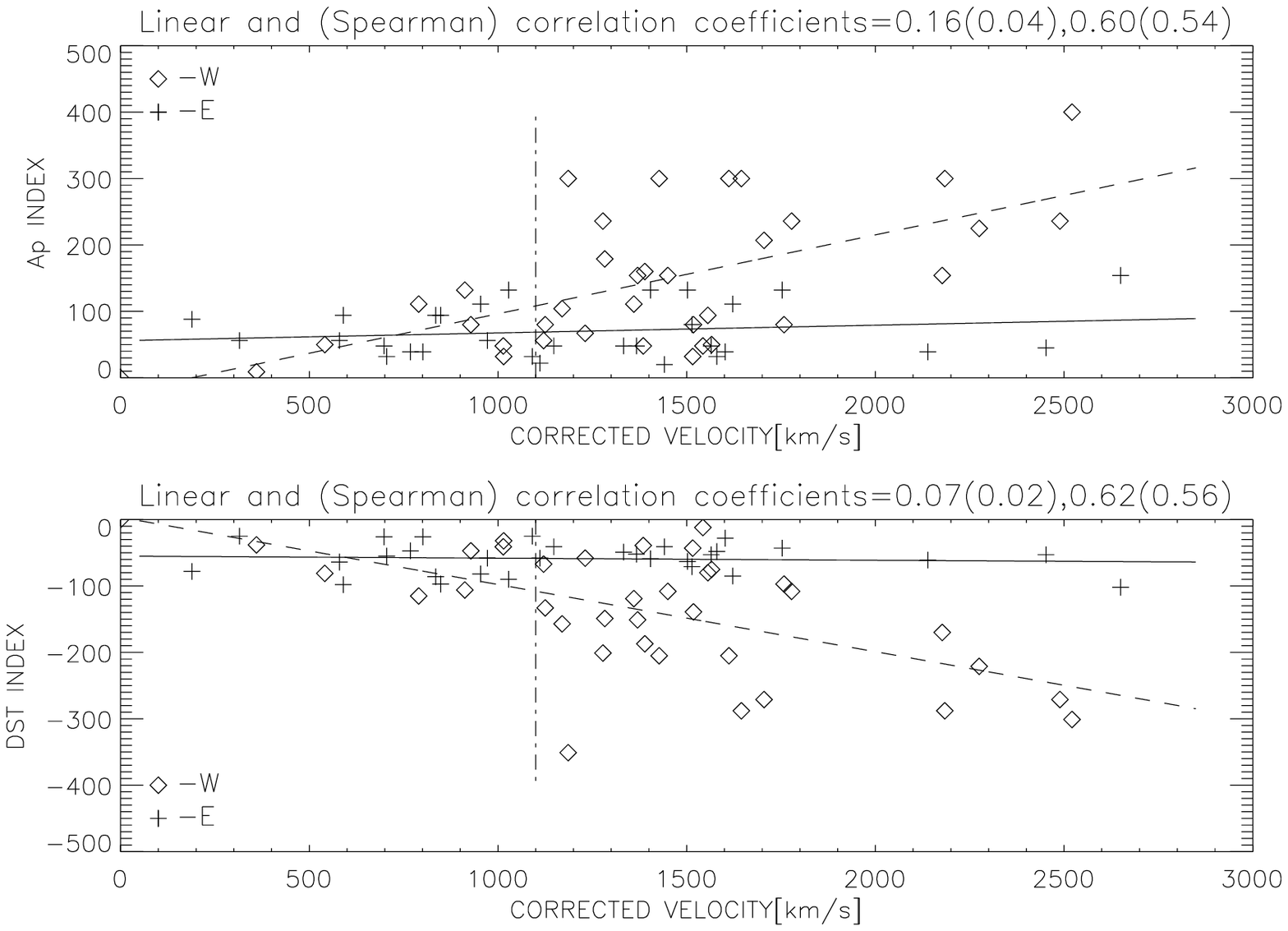} \vspace{0mm}\caption{The scatter plots of the space
velocity ($V$) versus $Ap$ and $D_{ST}$ indices. Diamond symbols
represent events originating from the western hemisphere and cross
symbols represent events originating from the eastern hemisphere.
The solid lines are the linear fits to  data points associated with,
eastern events and the dashed lines are linear fits to data points
associated with western events. The dot-dashed verticals lines
indicate velocity limits above which HCMEs can cause significant
geomagnetic storms ($D_{ST}\leq -150nT$). Upon inspection of this
 figure, it is clear that only very fast events ($V\geq1100km/s$) originating in the western
hemisphere can cause severe geomagnetic storms.}
\end{figure*}

\begin{figure*}
\vspace{9cm} \includegraphics{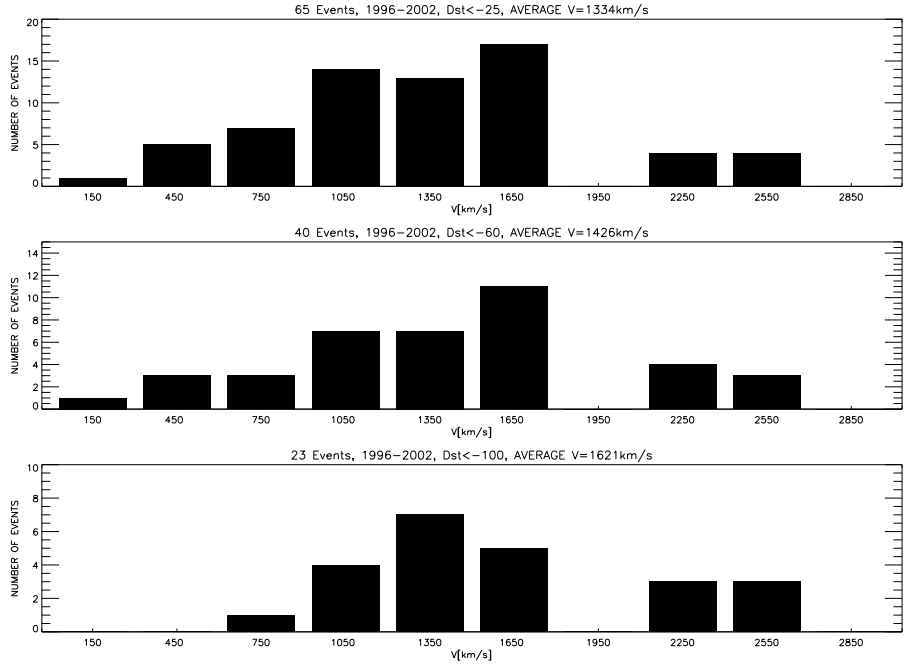} \vspace{0mm}\caption{The histograms showing
distribution of the space velocities ($V$) of FHCMEs which cause
geomagnetic disturbance with $D_{ST}$ index lower than $-25nT$,
$-60nT$ and $-100nT$. These histograms demonstrate  that
geoeffectivenes of HCMEs depend on their space velocities and sever
geomagnetic storms with $D_{ST}<-100nT$ can be caused by fast CMEs
(with $V>700km/s$) only.}
\end{figure*}

\begin{figure*}
\vspace{9cm} \includegraphics{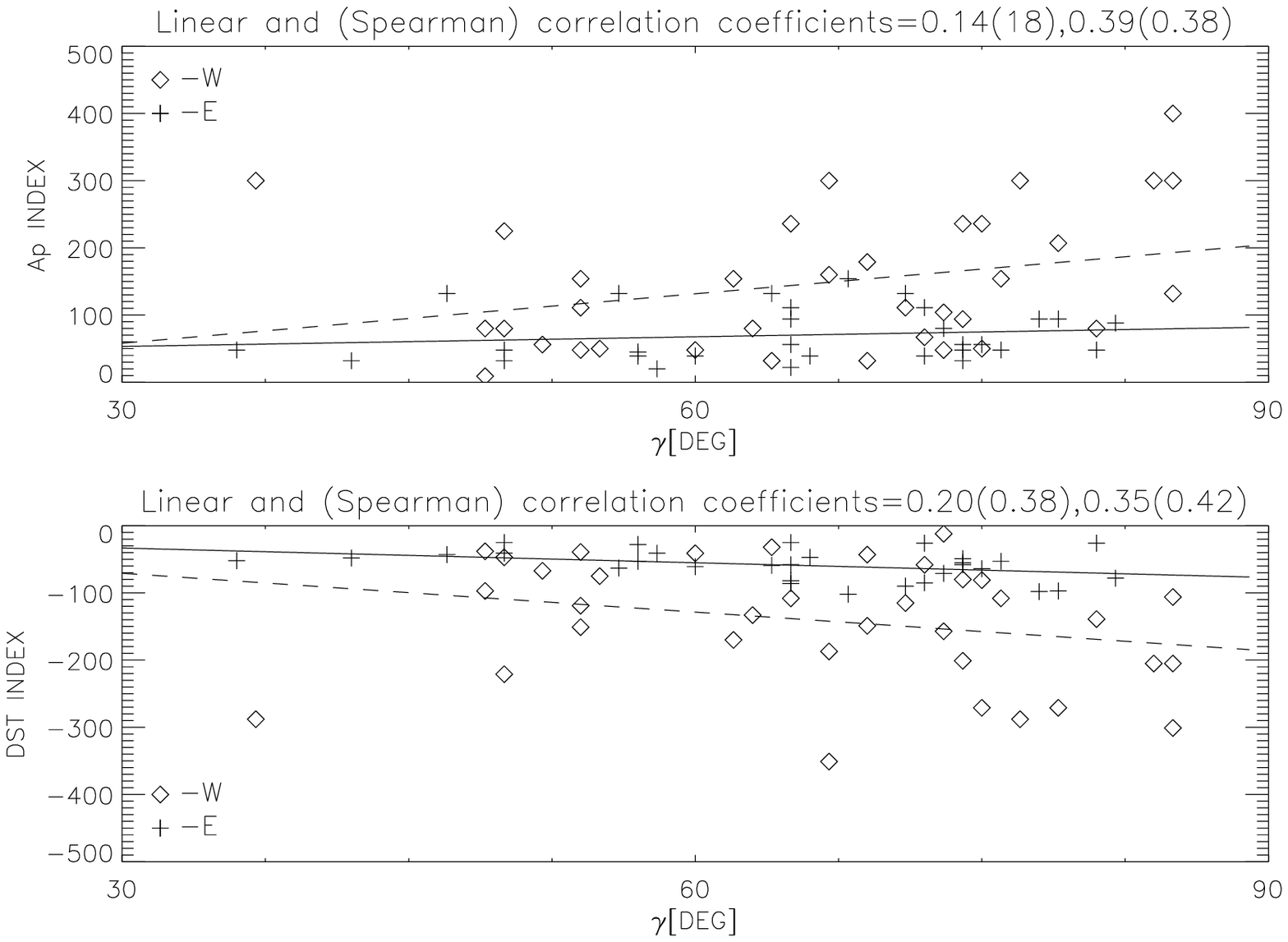} \vspace{0mm}\caption{The scatter plots of the source
location ($\gamma$) versus $Ap$ and $D_{ST}$ indices. Diamond
symbols represent events originating from the western hemisphere and
cross symbols represent events originating from the eastern
hemisphere. The solid lines are the linear fits to the data points
associated with eastern events, and the dashed lines are linear fits
to data points associated with western events. It is clear that the
western events originating close to the disk center
($\gamma\geq65^o$) are more likely to cause the biggest geomagnetic
storms. }
\end{figure*}

\begin{figure*}
\vspace{9cm} \includegraphics{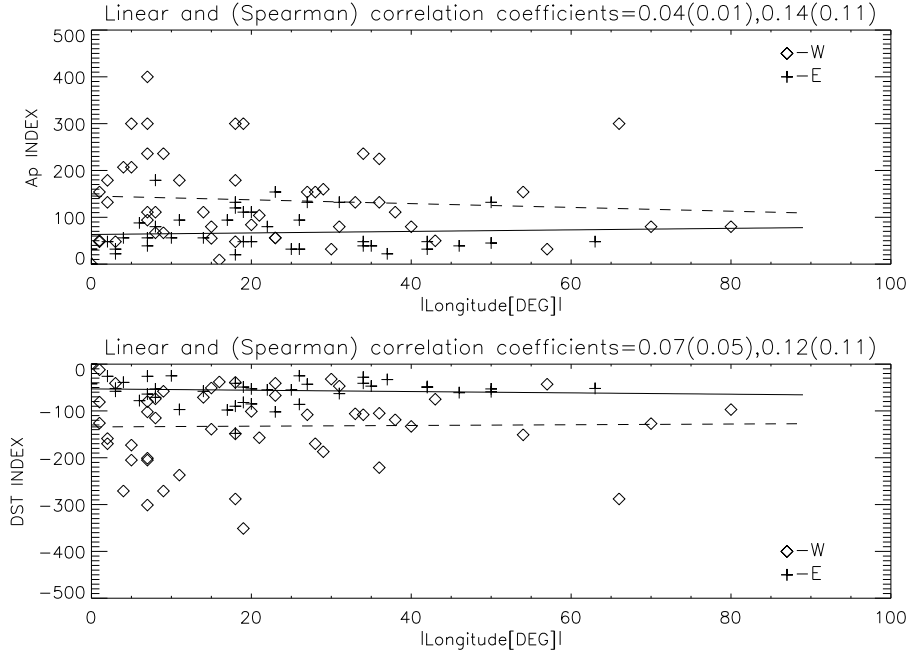} \vspace{0mm}\caption{The scatter plots of absolute
values of longitudes of H-alpha flares associated to HCMEs versus
$Ap$ and $D_{ST}$ indices. Diamond symbols represent events
originating from the western hemisphere and cross symbols represent
events originating from the eastern hemisphere. The solid lines are
the linear fits to the data points associated with eastern events,
and the dashed lines are linear fits to data points associated with
western events. Upon inspection of the figures, it is clear that the
western events originating close to the disk center  are more likely
to cause the biggest geomagnetic storms. }
\end{figure*}

\begin{figure*}
\vspace{9cm} \includegraphics{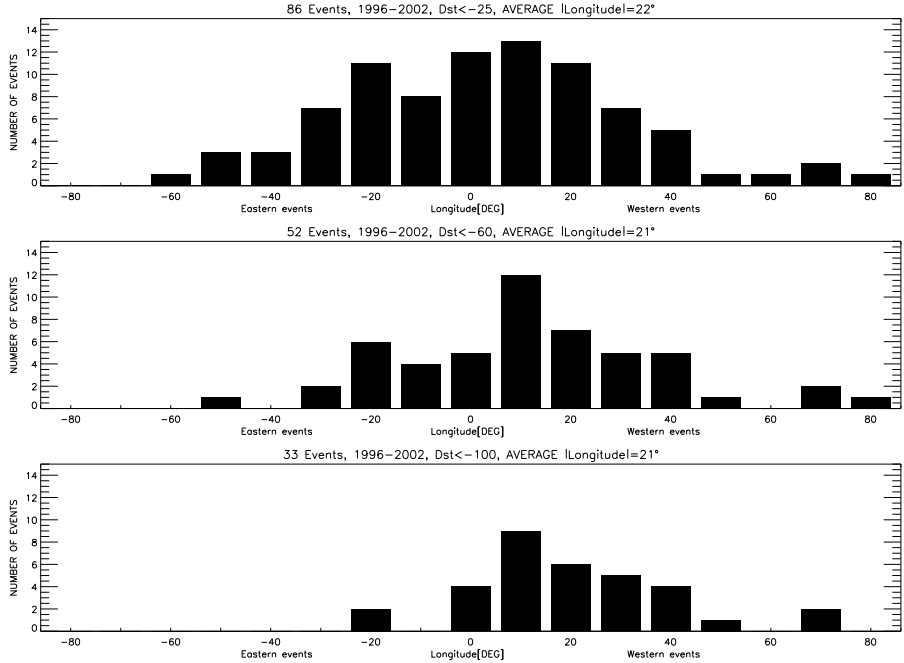} \vspace{0mm}\caption{The histograms showing
distribution of the longitude  of HCMEs which cause geomagnetic
disturbance with $D_{ST}$ index lower than $-25nT$, $-60nT$ and
$-100nT$. Upon inspection of the histograms, it is clear that the
goeffectiveness of CMEs depends on the longitude of source location
and that the severe geomagnetic disturbance ($D_{ST}<-100nT$) are
mostly caused by the western events originating close to the disk
center.}
\end{figure*}

\begin{figure*}
\vspace{9cm} \includegraphics{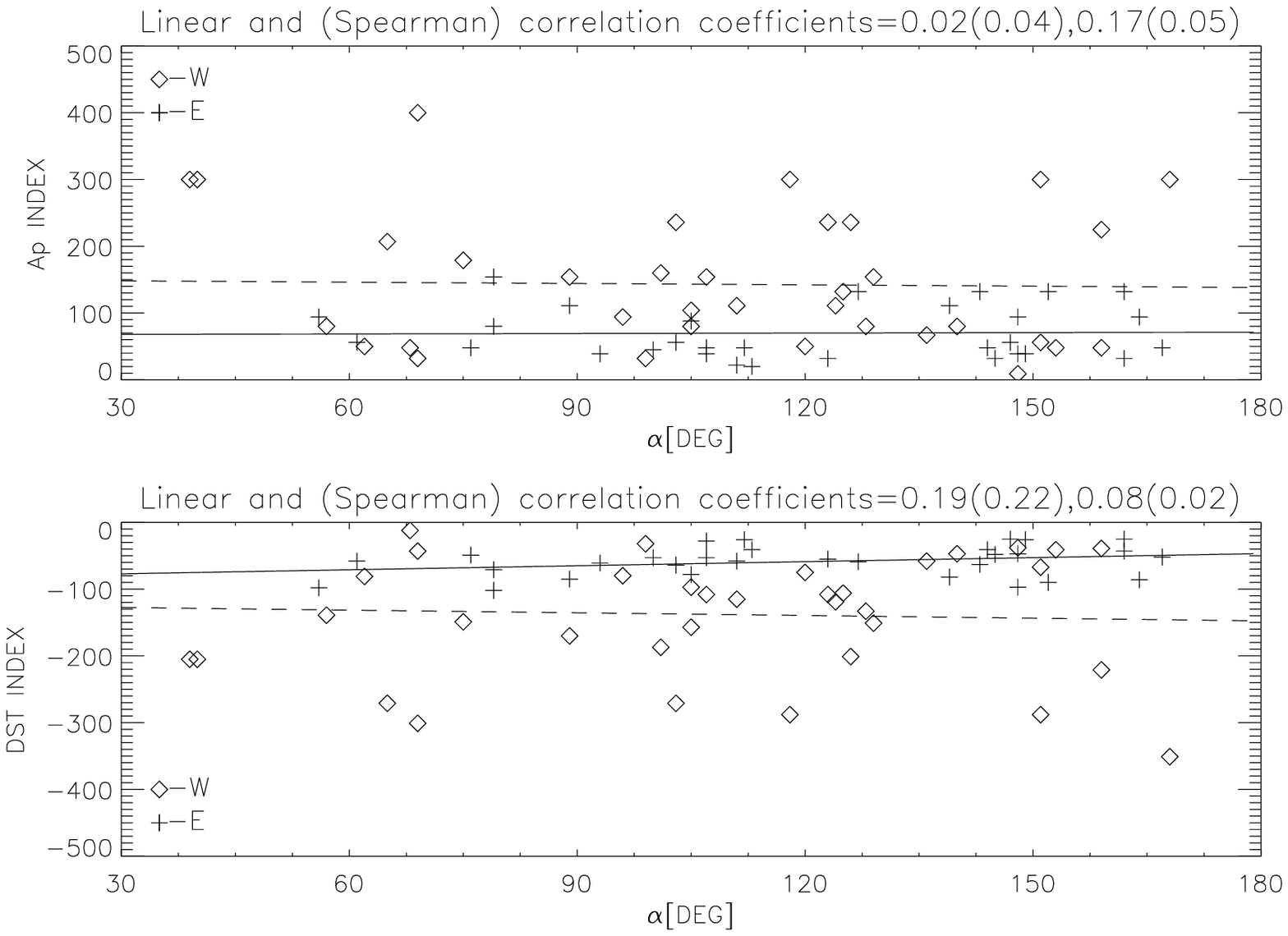} \vspace{0mm}\caption{The scatter plots of the $\alpha$
versus $Ap$ and $D_{ST}$ indices. Diamond symbols represent events
originating in the western hemisphere and cross symbols represent
events originating in the eastern hemisphere. The solid lines are
the linear fits to  data points associated with eastern events, and
the dashed lines are linear fits to data points associated with
western events.  The geoeffectiveness of CMEs depends very little on
their widths. }

\end{figure*}

\begin{figure*}
\vspace{11cm} \includegraphics{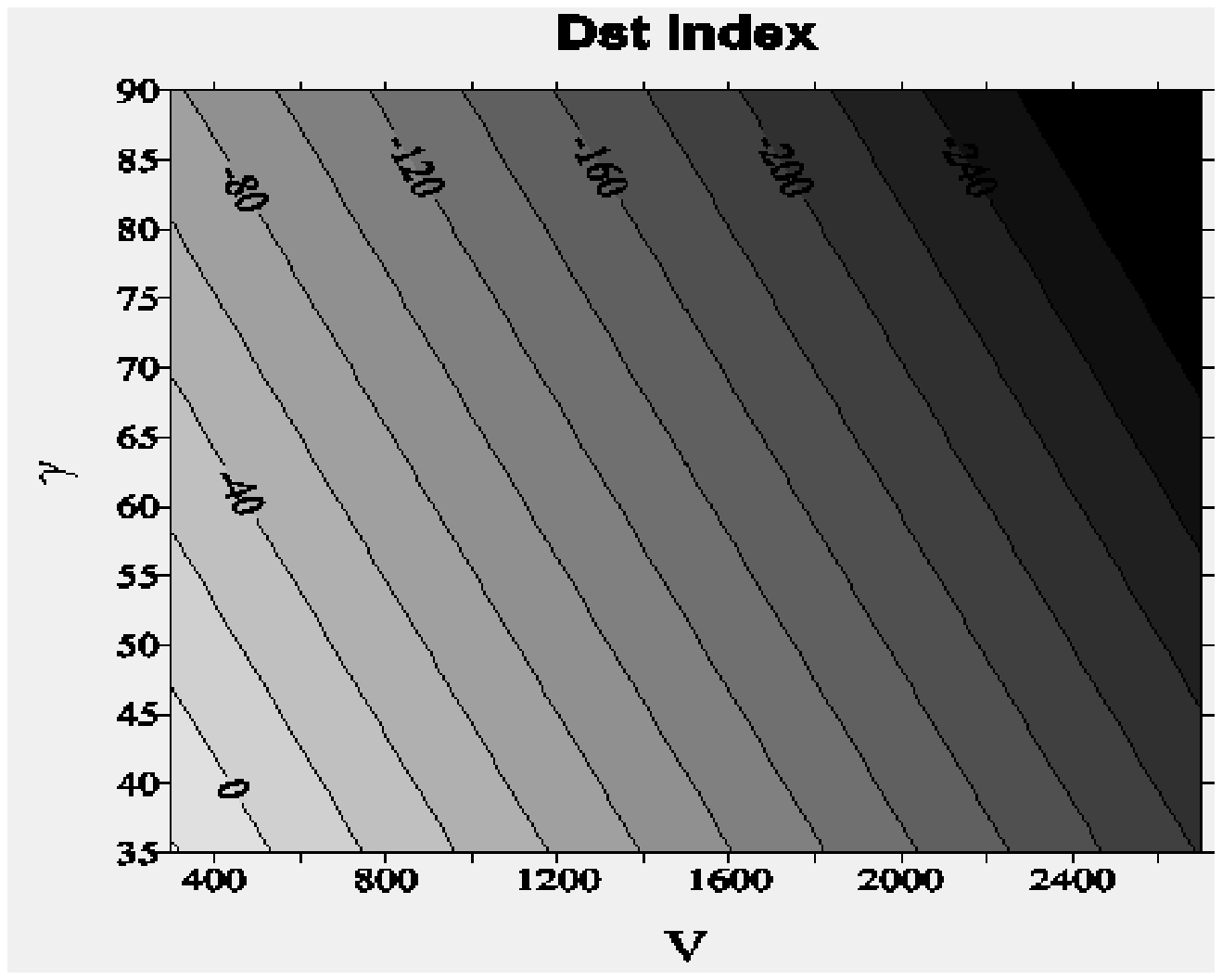} \vspace{0mm} \caption{The contour map
presenting
 $D_{ST}$ index versus the space velocity ($V$) and the source
 location ($\gamma$). From the inspection of the picture we see that the
strongest geomagnetic storms can occur for fast events originating
close to the disk center.}
\end{figure*}

\begin{figure*}
\vspace{11cm} \includegraphics{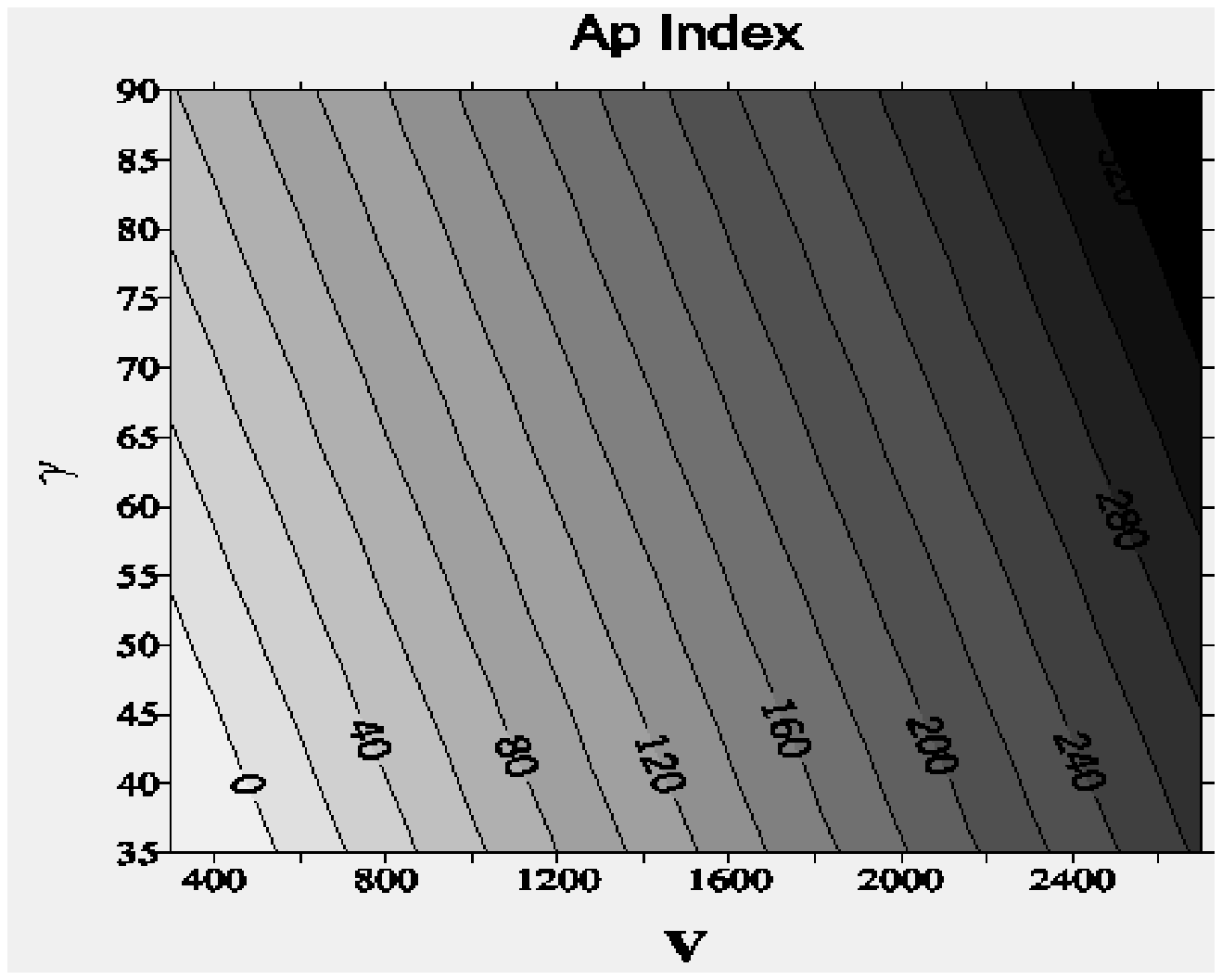} \vspace{0mm}\caption{The contour map
presenting
 $Ap$ index versus the space velocity ($V$) and the source
 location ($\gamma$). From the inspection of the picture we see that the
strongest geomagnetic storms can occur for fast events originating
close to the disk center.}
\end{figure*}

\begin{figure*}

\vspace{9cm} \includegraphics{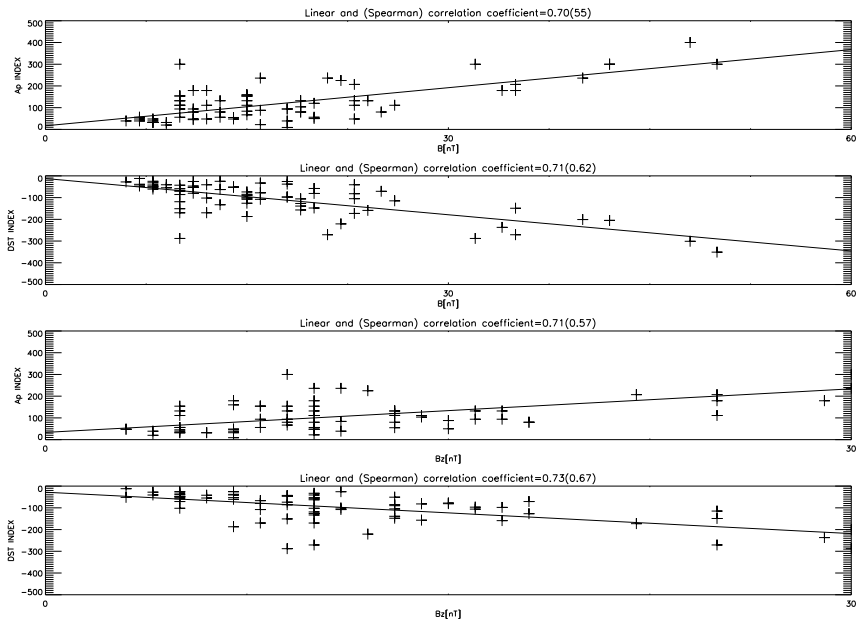} \vspace{0mm}\caption{The scatter plots of the $B$ and
$B_{Z}$ versus $Ap$ and $D_{ST}$ indices. Correlation between these
parameters and geomagnetic indices is significant (correlation
coefficient are $>0.50$) and linear and (Spearman) coefficients are
approximately equal 0.70(0.60) for ($B$) and ($B_Z$) as well.}
\end{figure*}

\begin{figure*}
\vspace{9cm} \includegraphics{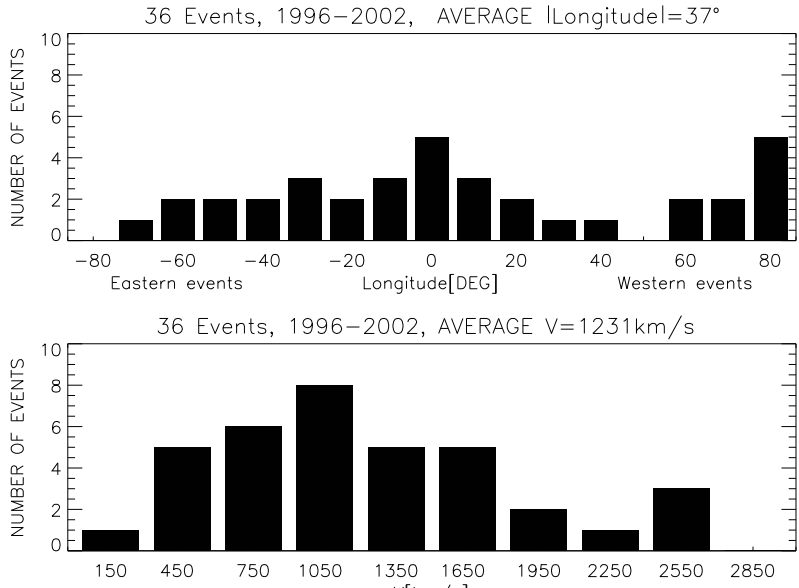} \vspace{0mm}\caption{The histogram showing
distributions of the longitude and space speed of non-geoeffective
FHCMEs. The histograms show that these events originate from the
whole solar disk and have velocities from $100km/s$ up to
$2500km/s$.}
\end{figure*}

\begin{figure*}
\vspace{9cm} \includegraphics{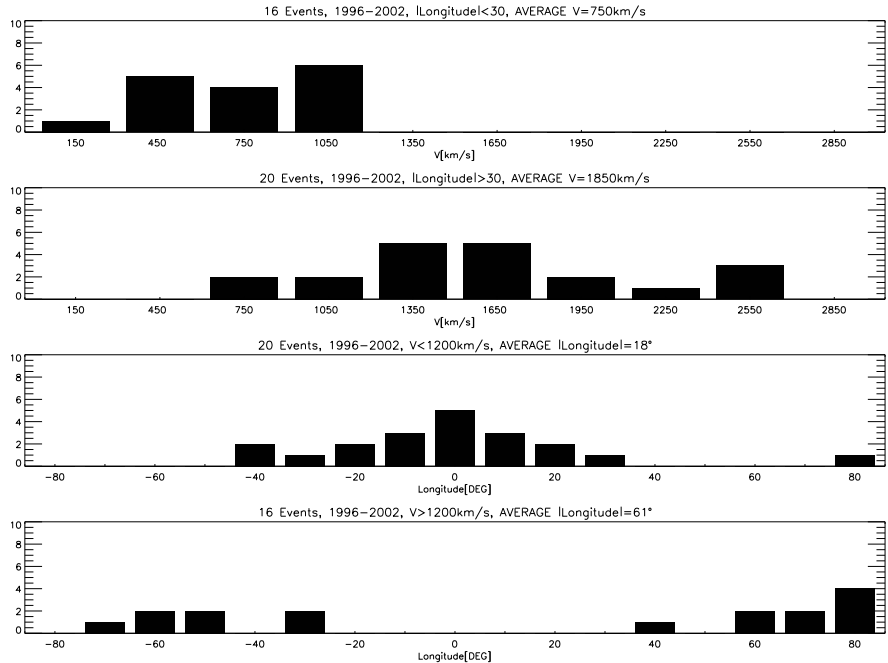} \vspace{0mm}\caption{The histograms showing: space
velocities for non-geoeffective FHCMEs originating close to the disk
center ($|longitude|<30^o$), space velocities for non-geoeffective
FHCMEs originating close to the limb ($|longitude|>30^o$), longitude
for slow non-geoeffective FHCMEs ($V<1200km/s$) and longitude for
fast non-geoeffective FHCME ($V>1200km/s$). Upon the inspection of
the histograms (the first and last panel in the figure) it is clear
that all fast HCMEs ($V>1200km/s$) originating close to the disk
center ($|longitud|<30^0$)  must be geoeffective.}
\end{figure*}

\begin{figure*}
\vspace{9cm} \includegraphics{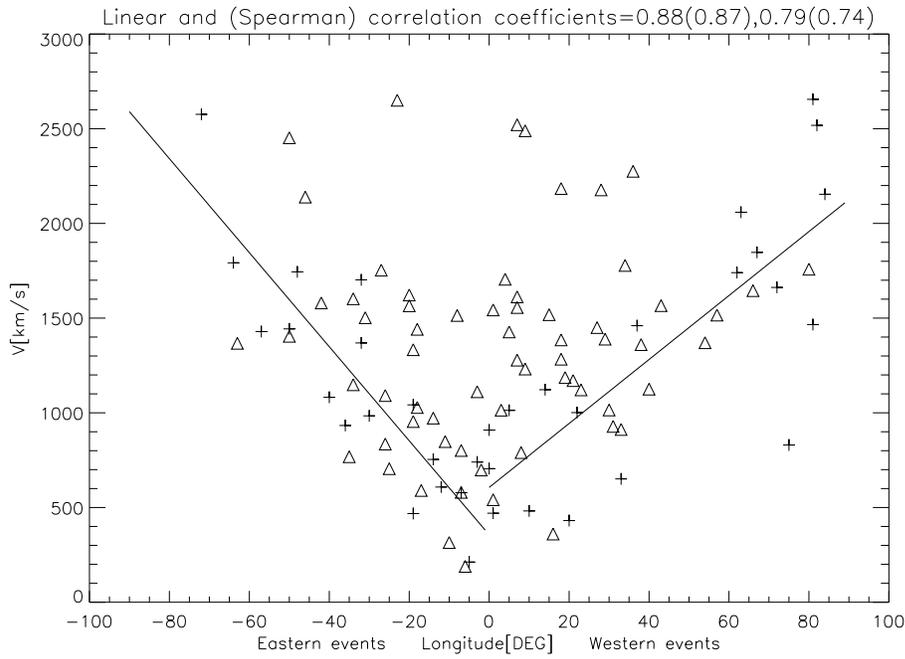} \vspace{0mm}\caption{Figure shows the scatter plot of
the space velocities versus longitude for all FHCMEs. Diamond
symbols represent goeeffective and cross symbols non-geoeffective
FHCMEs. The solid lines are linear fits for non-geoeffective events
originating from the east and west hemisphere.Upon inspection of the
figure, it is clear that geoeffective events are faster than the
 non-geoeffective events originating at the same
longitude.}
\end{figure*}

\newpage

\begin{table*}

\caption {The Table shows the list of frontside halo CMEs
(1996-2002). The
 first four columns are from the SOHO/LASCO catalog and give date, time of first appearance
 in the coronagraph field of view,
 projected speed and position angle of the fastest part of the HCME.
 Parameters $r,
\gamma, \alpha,$ and $V$, estimated from the cone model (Michalek et
al. 2003), are shown in columns  (5), (6), (7), and (8),
respectively. In column (9) the source locations of the associated
H-flares are presented. The changes of geomagnetic indices $D_{ST}$
and $Ap$ caused by the ICMEs are shown in columns 10 and 11. The
last two columns (12 and 13) give the maximum value of magnitude
($B$) and southward component ($B_Z$) of magnetic field in the
ICME.\\
\\
 \textbf{ List of halo CMEs}}


\begin{tabular}{c c c c c c c c c c c c c}
\hline  \hline
DATA & TIME & SPEED & PA & r & $\gamma$ & $\alpha$ & V & Location   &  Dst &  Ap  & B &  $B_{z}$ \\
\hline

  &    &   km/ s  &  Deg &
${1 \over R_{\odot}}$
&  Deg  & Deg
& km/s  &   &  &  & nT & nT \\
\hline


1996/12/02& 15:35:05&  538& 270&  0.77 & 39 & 110 &  830 & S05W75    & --- &---& ---  &--- \\
1997/01/06& 15:10:42&  136& 182&  0.13 & 82 & 105 &  189 & S18E06    & -78 & 88&  16  &-15\\
1997/04/07& 14:27:44&  875& 126&  0.42 & 65 & 139 &  954 & S30E19    & -82 &111&  23  &-14 \\
1997/05/12& 06:30:09&  464& 277&  0.32 & 71 & 111 &  790 & N21W08    &-115 &111&  26  &-25 \\
1997/08/30& 01:30:35&  370& 65 &  0.21 & 78 &  56 &  590 & N30E17    & -98 & 94&  18  &-17\\
1997/09/17& 20:28:48&  377& 274&  0.65 & 49 & 148 &  360 & N45W16    & -38 & 09&  18  &-07  \\
1997/09/28& 01:08:33&  359&  66&  0.53 & 57 & 131 &  212 & N22E05    &---  &---& ---  &---  \\
1997/10/21& 18:03:45&  523&  30&  0.24 & 75 & 103 &  580 & N16E07    & -64 & 56& ---  &--- \\
1997/11/04& 06:10:05&  755& 265&  0.12 & 85 & 125 &  912 & S14W33    &-106 &132&  19  &-16 \\
1997/11/06& 12:10:41& 1556& 261&  0.82 & 34 & 153 & 2059 & S18W63    &---  &---& ---  &---\\
1998/01/21& 06:37:25&  361& 176&  0.71 & 44 & 159 &  468 & S57E19    &---  &---& ---  &--- \\
1998/01/25& 15:26:34&  693&  36&  0.27 & 74 & 123 &  705 & N21E25    & -55 & 32&  09  &-06  \\
1998/04/23& 05:27:07& 1618& 113&  0.61 & 59 & 126 & 1744 & E15E48    &---  &---& ---  &---\\
1998/04/27& 08:56:06& 1385&  79&  0.35 & 69 & 118 & 1443 & S16E50    &---  &---& ---  &---\\
1998/04/29& 16:58:54& 1374&  16&  0.30 & 72 &  89 & 1622 & S17E20    & -85 &111&  15  &-13 \\
1998/05/01& 23:40:09&  585& 142&  0.10 & 84 &  40 & 1427 & S18W05    &-205 &300&  42  &-35 \\
1998/05/02& 05:31:56&  542& 143&  0.10 & 85 &  39 & 1612 & S20W07    &-205 &300&  42  &-35 \\
1998/05/02& 14:06:12&  938& 311&  0.15 & 81 &  57 & 1518 & S15W15    &-139 & 80&  19  &-13  \\
1998/11/04& 07:54:07&  523&   0&  0.25 & 75 &  62 &  541 & N17W01    & -81 & 50&  20  &-15\\
1998/11/05& 02:02:52&  380& 264&  0.18 & 79 &  88 &  482 & N19W10    &---  &---& ---  &---\\
1998/11/05& 20:44:02& 1118& 305&  0.35 & 69 &  75 & 1283 & N22W18    &-149 &179&  35  &-25 \\
1998/11/24& 02:30:05& 1744& 224&  0.88 & 27 & 153 & 2655 & S30W81    &---  &---& ---  &---\\
1998/11/27& 08:30:05&  434& ---&   --- &--- & --- & ---  & S24E09    &---  &---& ---  &---\\
1998/12/18& 18:09:50& 1749&  40&  0.68 & 47 & 120 & 1792 & N19E64    &---  &---& ---  &---\\
1999/05/03& 06:06:05& 1584&  50&  0.61 & 51 & 110 & 1369 & N15E32    &---  &---& ---  &---\\
1999/05/10& 05:50:05&  920&  80&  0.27 & 74 &  76 & 1333 & N16E19    &-49  & 48& ---  &---  \\
1999/06/12& 21:26:08&  465& ---&  ---  & ---& --- &  --- & N27W43    & --- &---& ---  &---\\
1999/06/22& 18:54:05& 1133& ---&  ---  & ---& --- &  --- & N22E37    &-33  & 22&  16  &-10\\
1999/06/23& 07:31:24& 1006& ---&  ---  & ---&  ---&  --- & N24E45    &---  &---& ---  &---\\
1999/06/24& 13:31:24&  975& 314&  0.63 & 50 & 144 & 1148 & N24E34    &-41  & 48&  12  &-07\\
1999/06/26& 07:31:25&  558&   0&  0.11 & 83 &  67 &  909 & N25E00    &---  &---& ---  &---  \\
1999/06/28& 21:30:08& 1083& ---&  ---  & ---& --- &  --- & N26W51    &---  &---& ---  &--- \\
1999/06/29& 07:31:26&  634&  10&  0.15 & 81 & 112 &  698 & N19E02    & -26 & 48&  11  &-07  \\
1999/06/29& 18:54:07&  438& ---&  ---  & ---& --- &  --- & S14E01    & --- &---& ---  &---\\
1999/06/30& 11:54:07&  406&  23&  0.16 & 80 &  92 &  705 & S15E00    & --- &---& ---  &---  \\
1999/07/25& 13:31:21& 1389& 306&  0.76 & 40 & 127 & 1466 & N38W81    & --- &---& ---  &---  \\
1999/07/28& 05:30:05&  457& ---&  ---  & ---& --- &  --- & S15E08    &-53  &179&  11  &-07  \\
1999/07/28& 09:06:05&  456& ---&  ---  & ---& --- &  --- & S15E04    &-39  &56 &  07  &-05  \\
1999/09/20& 06:06:05&  604& ---&  ---  & ---& --- &  --- & S20W05    &-173 &207&  23  &-22  \\
1999/10/14& 09:26:05& 1250&  63&  0.82 & 34 & 157 & 1702 & N11E32    &---  &---& ---  &---\\
1999/12/22& 02:30:05&  570&  14&  0.75 & 40 & 162 &  984 & N10E30    &---  &---& ---  &---\\
1999/12/22& 19:31:22&  605&  24&  0.65 & 69 & 141 & 1042 & N24E19    &---  &---& ---  &---\\
2000/01/18& 17:54:05&  739& 162&  0.18 & 79 & 148 &  848 & S19E11    &-97  & 94&  18  &-16     \\
2000/01/28& 20:12:41& 1177& ---&  ---  & ---& --- &  --- & S31W28    &---  &---& ---  &---   \\
2000/02/08& 09:30:05& 1079&  55&  0.63 & 50 & 162 & 1091 & N25E26    &-25  & 32&  08  &-07    \\
2000/02/09& 19:54:17&  910& 218&  0.44 & 63 & 128 & 1125 & S17W40    &-133 & 80&  13  &-10  \\
2000/02/10& 02:30:05&  944& 331&  0.41 & 65 & 111 & 1111 & N30E03    & -58 & 22& ---  &---   \\
2000/02/12& 04:31:20& 1107& 335&  0.61 & 52 & 151 & 1121 & N26W23    & -67 & 56&  10  &-08   \\
2000/02/17& 20:06:05&  728& 196&  0.29 & 72 & 149 &  801 & S29E07    & -26 & 39&  18  &-11 \\
2000/04/04& 16:32:37& 1188& 304&  0.79 & 37 & 151 & 1645 & N16W66    &-288 &300&  32  &-30  \\
2000/04/10& 00:30:05&  409& 212&  0.14 & 81 & 125 &  470 & S14W01    & --- &---&  --- &---    \\
2000/05/05& 15:50:05& 1594& 269&  0.85 & 32 & 146 & 2154 & S16W84    & --- &---&  --- &---   \\
2000/06/06& 15:54:05& 1119&   6&  0.32 & 71 & 152 & 1028 & N20E18    & -90 &132&   15 &-13  \\
2000/06/07& 16:30:05&  842& ---&  ---  & ---& --- &  --- & N23E03    & -52 & 32&  --- &---   \\
2000/06/10& 17:08:05& 1108& 306&  0.68 & 50 & 138 & 1460 & N22W37    & --- &---&  --- &---   \\
2000/07/07& 10:26:05&  453& 198&  0.42 & 65 & 147 &  315 & N17E10    & -25 & 56&   13 &-05  \\
2000/07/11& 13:27:23& 1078&  51&  0.68 & 47 & 162 & 1753 & N18E27    & -43 &132&   10 &-09    \\
2000/07/14& 10:54:07& 1674& 270&  0.10 & 85 &  69 & 2521 & N22W07    &-301 &400&   48 &-40  \\
2000/07/25& 03:30:05&  528& ---&  ---  &--- & --- &  --- & N06W08    & -74 &67 &   15 &-09  \\
2000/08/09& 16:30:05&  702& ---&  ---  & ---& --- &  --- & N11W11    &-237 &179&   34 &-29\\
2000/09/12& 11:54:05& 1550& 216&  0.58 & 54 & 159 & 1385 & S17W18    & -39 &48 &  --- &---  \\
2000/09/15& 21:50:07&  257& ---&  ---  & ---& --- &  --- & N13E03    & --- &---&  --- &--- \\
2000/09/16& 05:18:14& 1251&  21&  0.27 & 74 & 126 & 1278 & N14W07    &-201 &236&   40 &-30\\
2000/09/25& 02:50:05&  587& ---&  ---  & ---& --- &  --- & S11W59    &---  &---&  --- &--- \\
2000/10/02& 03:50:05&  525& 144&  0.41 & 65 & 131 &  578 & S09E07    &---  &---&  --- &--- \\
2000/10/02& 20:26:05&  569& ---&  ---  & ---& --- &  --- & S10W02    &-170 &179&   12 &-10  \\
\hline
\end{tabular}
\end{table*}

\newpage
\begin{table*}
\caption{List of halo CMEs}

\begin{tabular}{c c c c c c c c c c c c c}
\hline \hline
DATA & TIME & SPEED & PA & r & $\gamma$ & $\alpha$ & V & Location   &  Dst &  Ap  & B &  $B_{z}$ \\
\hline

 &    &   km/ s  &  Deg &
${1 \over R_{\odot}}$ &  Deg  & Deg
& km/s  &   &  &  & nT & nT \\
\hline

2000/10/09& 23:50:05&  798& ---&  ---  & ---& --- &  --- & N01W14    & -71 &111&   10 &-05  \\
2000/10/24& 08:26:05&  800& ---&  ---  & ---& --- &  --- & S20E70    &---  &---&  --- &---  \\
2000/10/25& 08:26:05&  770& ---&  ---  & ---& --- &  --- & N17W70    &-127 &80 &   19 &-18  \\
2000/11/01& 16:26:08&  801& ---&  ---  & ---& --- &  --- & S20E42    & -50 &48 &  --- &--- \\
2000/11/03& 18:26:06&  291& ---&  ---  & ---& --- &  --- & N02W02    &-159 &132&   24 &-17  \\
2000/11/23& 06:06:05&  492& 230&  0.72 & 43 & 168 &  651 & S22W33    &---  &---&  --- &--- \\
2000/11/24& 05:30:05&  994& 352&  0.45 & 62 & 147 & 1013 & N20W05    &---  &---&  --- &---  \\
2000/11/24& 15:30:05& 1245& 324&  0.26 & 74 &  96 & 1556 & N22W07    & -80 &94 &   11 &-08  \\
2000/11/24& 22:06:05& 1005& 312&  0.56 & 55 & 130 & 1122 & N21W14    &---  &---&  --- &--- \\
2000/11/25& 01:31:58& 2519&  75&  0.54 & 57 & 100 & 2452 & N07E50    &-53  &45 &   11 &-05  \\
2000/11/25& 09:30:17&  675& ---&  ---  & ---& --- &  --- & N18W24    &---  &---&  --- &---   \\
2000/11/25& 19:31:57&  671& ---&  ---  & ---& --- &  --- & N20W23    &---  &---&  --- &---  \\
2000/11/26& 17:06:05&  980& 283&  0.58 & 54 & 124 & 1360 & N18W38    &-119 &111&   10 &-10\\
2000/12/18& 11:50:05&  510&  15&  0.05 & 87 & 105 &  740 & N14E03    &---  &---&  --- &---    \\
2001/01/10& 00:54:05&  832& 102&  0.55 & 56 & 122 &  933 & N13E36    &---  &---&   10 &-05\\
2001/01/20& 19:31:50&  839& 109&  0.51 & 58 & 134 & 1082 & S07E40    &---  &---&  --- &---    \\
2001/01/20& 21:30:08& 1507&  78&  0.49 & 60 &  93 & 2139 & S07E46    & -61 &39 &   08 &-07  \\
2001/02/11& 01:31:48& 1183& 294&  0.64 & 69 &  69 & 1516 & N24W57    & -43 &32 &  --- &--- \\
2001/02/15& 13:54:05&  625&  18&  0.39 & 66 & 134 &  608 & N07E12    & --- &---&   07 &-06\\
2001/03/19& 05:26:05&  389& ---&  ---  & ---& --- &  --- & S05W00    & -42 &32 &   08 &-06 \\
2001/03/24& 20:50:05&  906& ---&  ---  & ---& --- &  --- & N15E22    & -55 &80 &  --- &---  \\
2001/03/25& 17:06:05&  677&  19&  0.42 & 65 & 164 &  835 & N16E26    & -86 &94 &   10 &-09  \\
2001/03/29& 10:26:05&  942& 277&  0.38 & 67 & 168 & 1186 & N20W19    &-351 &300&   50 &-45\\
2001/04/05& 17:06:05& 1390&  83&  0.43 & 64 & 127 & 1404 & S24E50    & -59 &132&   10 &-05  \\
2001/04/06& 19:30:02& 1270& 124&  0.54 & 56 & 143 & 1502 & S21E31    & -63 &132&   13 &-10 \\
2001/04/09& 15:54:02& 1192& 213&  0.18 & 79 &  65 & 1705 & S21W04    &-271 &207&   35 &-25   \\
2001/04/10& 05:30:00& 2411& 187&  0.25 & 75 & 103 & 2489 & S23W09    &-271 &236&   21 &-10\\
2001/04/11& 13:31:48& 1103& 229&  0.23 & 76 & 107 & 1450 & S22W27    &-108 &154&   15 &-08 \\
2001/04/12& 10:31:29& 1184& 251&  0.57 & 55 & 120 & 1566 & S19W43    & -75 & 50&  --- &---  \\
2001/04/26& 12:30:05& 1006&  46&  0.64 & 50 & 140 &  928 & N17W31    & -47 & 80&   11 &-09 \\
2001/08/14& 16:01:28&  618& ---&  ---  & ---& --- &  --- & N16W36    &-105 &132&   23 &-13  \\
2001/08/25& 16:50:05& 1433& 127&  0.54 & 57 & 107 & 1602 & S17E34    & -28 & 39&   06 &-04 \\
2001/08/31& 16:11:33&  310& ---&  ---  & ---& --- &  --- & N19E33    & --- &---&  --- &---   \\
2001/09/11& 14:54:05&  791&  74&  0.39 & 66 & 148 & 768  & N13E35    & -47 & 39&   07 &-05\\
2001/09/24& 10:30:59& 2402& 147&  0.36 & 68 &  79 & 2650 & S12E23    &-102 &154&   15 &-05\\
2001/09/28& 08:54:34&  846& ---&  ---  & ---&  ---&  --- & N10E18    &-148 &120&   20 &-13 \\
2001/10/09& 11:30:05&  973& 174&  0.27 & 73 &  79 & 1514 & S28E08    & -71 & 80&   25 &-18 \\
2001/10/19& 16:50:05&  901& 268&  0.37 & 67 & 101 & 1389 & N15W29    &-187 &160&   15 &-07 \\
2001/10/22& 15:06:05& 1336& 121&  0.51 & 58 & 113 & 1441 & S21E18    & -41 & 20&   09 &-04 \\
2001/10/25& 15:26:05& 1092& 203&  0.27 & 73 & 105 & 1170 & S16W21    &-157 &104&   19 &-14 \\
2001/11/01& 22:30:05&  453& ---&  ---  & ---& --- &  --- & N12W23    & -41 & 56&  --- &---  \\
2001/11/03& 19:20:05&  457& 307&  0.25 & 75 & 138 &  431 & N04W20    &---  &---&  --- &--- \\
2001/11/04& 16:35:06& 1810& 246&  0.21 & 77 & 118 & 2184 & N06W18    &-288 &300&   10 &-09\\
2001/11/17& 05:30:06& 1379&  65&  0.73 & 42 & 145 & 1580 & S13E42    & -48 & 32&   08 &-05 \\
2001/11/21& 14:06:05&  518& ---&  ---  &--- & --- &  --- & S13W18    & --- &---&  --- &--- \\
2001/11/22& 20:30:33& 1443& 221&  0.74 & 42 & 138 & 1847 & S25W67    & --- &---&  --- &--- \\
2001/11/22& 23:30:05& 1437&  68&  0.74 & 42 & 159 & 2275 & S17W36    &-221 &225&   22 &-12\\
2001/11/28& 17:30:06&  500& ---&   --- & ---& --- & ---  & N14E16    &---  &---&   07 &-03 \\
2001/12/13& 14:54:06&  864& ---&  ---  &--- & --- & ---  & N16E09    &---  &---&  --- &---\\
2002/01/04& 09:30:05&  896& ---&  ---  & ---& --- &  --- & N38E87    &---  &---&  --- &---\\
2002/01/08& 17:54:05& 1794& ---&  ---  & ---& --- &  --- & S18E79    & -51 & 48&  --- &--- \\
2002/02/20& 06:30:05&  952& 270&  0.88 & 27 & 153 & 1662 & N12W72    &---  &---&  --- &--- \\
2002/03/14& 17:06:06&  907& 140&  0.75 & 41 & 132 & 1429 & S23E57    &---  &---&  --- &--- \\
2002/03/15& 23:06:06&  907& 276&  0.49 & 60 & 153 & 1014 & S08W03    &-41  &48 &   23 &-10\\
2002/03/18& 02:54:06&  989& 257&  0.31 & 71 & 115 & 1001 & S15W22    &---  &---&   08 &-05 \\
2002/03/20& 17:30:05&  308& ---&  ---  & ---& --- & ---  & S17W20    &-101 & 84&   15 &-11\\
2002/03/22& 11:06:05& 1750& 261&  0.64 & 49 & 105 & 1758 & S20W80    & -97 & 80&  --- &---\\
2002/04/15& 03:50:05&  720& ---&  ---  &--- & --- &  --- & S15W01    &-126 &154&   15 &-10\\
2002/04/17& 08:26:05& 1218& 283&  0.58 & 54 & 129 & 1370 & S16W54
&-151 &154&   10 &-09\\2
2002/05/07& 04:06:05&  720& 120&  0.32 & 70 &  77 &  754 & S22E14    &---  &---&   08 &-03\\
2002/05/08& 13:50:05&  614& ---&  ---  &--- & --- &  --- & S12W07    &-102 &111&   12 &-10\\
2002/05/16& 00:50:05&  600& 140&  0.26 & 74 &  61 &  972 & S22E14    & -58 & 56&   20 &-10\\
2002/05/22& 03:50:05& 1494& 247&  0.45 & 65 & 123 & 1778 & S30W34    &-108 &236&   16 &-11\\
2002/05/28& 16:26:05& 1244& 224&  0.89 & 26 & 158 & 2518 & N06W82    &---  &---&  --- &--- \\
2002/07/15& 20:30:05& 1132&   3&  0.28 & 73 &  68 & 1543 & N19W01    & -12 & 48&   07 &-03\\
2002/07/18& 08:06:08& 1111& 279&  0.43 & 64 &  99 & 1015 & N19W30    & -32 & 32&   08 &-05 \\
2002/07/23& 01:31:51& 1726&  87&  0.87 & 29 & 141 & 2576 & S13E72    &---  &---&  --- &---\\
2002/07/26& 22:06:10&  818& ---&  ---  & ---& --- &  --- & S19E26    &---  &---&  --- &---\\
2002/07/29& 12:07:33&  556& ---&  ---  & ---& --- &  --- & N10W15    &-51  &55 &   14 &-13 \\
2002/08/16& 12:30:05& 1459& 121&  0.23 & 76 & 107 & 1565 & S14E20    &-53  &48 &   14 &-10 \\
2002/08/22& 02:06:06& 1005& 233&  0.77 & 39 & 139 & 1740 & S05W62    &---  &---&  --- &---\\
2002/08/24& 01:27:19& 1878& ---&  ---  &--- & --- &  --- & S02E81    &-45  &39 &  --- & ---\\
2002/09/05& 16:54:06& 1657& 115&  0.46 & 62 &  89 & 2177 & N04W28    &-170 &154&   10 &-08 \\
2002/11/09& 13:31:45& 1838& ---&  ---  & ---& --- &  --- & S12W29    &---  &---&  --- &---\\
2002/11/10& 03:30:11& 1516& ---&  ---  & ---& --- &  --- & S12W37    &---  &---&  --- &---\\
2002/11/24& 20:30:05& 1077&  48&  0.79 & 36 & 167 & 1367 & N15E63    &-52  & 48&   08 &-03\\
2002/12/08& 23:54:05& 1339& ---&  ---  & ---& --- &  --- & S18E70    &---  &---&  --- &---\\
2002/12/19& 22:06:05& 1092& 299& 0.30  & 72 & 136 & 1231 & N15W09    &-58  &67 &  --- &---\\
\hline
\end{tabular}
\end{table*}
\end{article}
\end{document}